\documentstyle[twoside,fleqn,espcrc2,psfig]{article}

\newcommand{\AmS}{{\protect\the\textfont2
  A\kern-.1667em\lower.5ex\hbox{M}\kern-.125emS}}

\title{BeppoSAX Observations of the Seyfert 1 Galaxy NGC 3516}

\author{G.M. Stirpe\address{Osservatorio Astronomico di
        Bologna,\\ 
        Via Zamboni 33, 40126 Bologna, Italy},
        B.J. Wilkes\address{Harvard-Smithsonian Center for
        Astrophysics, \\
        60 Garden Street, Cambridge MA 02138, USA},
        A. Comastri$^{\rm a}$, S. Mathur$^{\rm b}$,
        P.T. O'Brien\address{Department of Physics \& Astronomy, University of
        Leicester,\\
        University Road, Leicester, UK}}
       
\begin{document}

\begin{abstract} We present the results of two observations of the bright
Seyfert 1 galaxy NGC 3516, obtained with BeppoSAX in 1996 November and 1997
March.  Useful signal is detected between 0.2 and 60~keV, allowing for the first
time the simultaneous observation of all main spectral features.  The source was
brighter by a factor 2 at the second epoch of observation.  Both spectra present
a strong Fe K$\alpha$ line, and a reflection hump at high energy.  An absorption
edge at 0.8~keV is visible in the later spectrum, but not in the earlier one,
indicating that this feature is strongly variable.  \end{abstract}

\maketitle

\section{INTRODUCTION}

The bright Seyfert 1 galaxy NGC 3516 was noted in the '80s for the detection in
IUE spectra of a broad ($\sim2000$~km~s$^{-1}$) trough cutting into the
C~IV~$\lambda$1549 emission line at $\sim -500$~km~s$^{-1}$ with respect to the
line peak (\cite{Koratkar96}, and references therein).  This feature,
attributed to line-of-sight absorption within the source itself, had disappeared
by 1993, and has not been detected again.  Similar features were observed in the
optical Balmer lines during the LAG monitoring campaign of 1990 
\cite{Wanders93}.

More recent X-ray data from Ginga, ROSAT, and ASCA have shown that NGC 3516 has
a warm absorber (O~VII/O~VIII and Fe edges have been detected), as reported by
\cite{Kolman93,Kriss96,Mathur97}.  The
latter show that the properties of the O~VII/O~VIII edges are compatible with
N$_H\sim 7\times10^{21}$~cm$^{-2}$ and ionization factor $U\sim 8$--13.  The
warm absorber is also variable.

In \cite{Mathur97} it is suggested that the absorption features in the
UV/optical and X-ray bands are all due to the same gas component as it evolves
in time:  an outflow and expansion compatible with the kinematic characteristics
of the UV/optical absorption would produce the increase in ionization required
to explain the disappearance of the UV/optical absorption itself, while the
X-ray warm absorber would be still visible.  The prediction is that the warm
absorber should also disappear on a time-scale of a few years as the ionization
of the gas increases further.

We have therefore observed NGC 3516 with all narrow-field instruments of
BeppoSAX at 2 epochs (1996 November and 1997 March, 50 ksec per observation) in
order to test this prediction.  In addition we have HST spectra from 5 epochs of
1996/1997, taken to check whether the broad absorption should reappear in the UV
band.

Further aims of the project are (a) to use the wide spectral range of BeppoSAX 
to study the broad-band continuum characteristics such as the reflection 
component at high energy, and (b) to study the Fe K$\alpha$ line first detected 
by ASCA \cite{Nandra97}.

This paper presents preliminary results from the BeppoSAX observations.

\section{BROAD BAND CHARACTERISTICS}
\begin{figure*}
\begin{minipage}[b]{.46\linewidth}
\psfig{figure=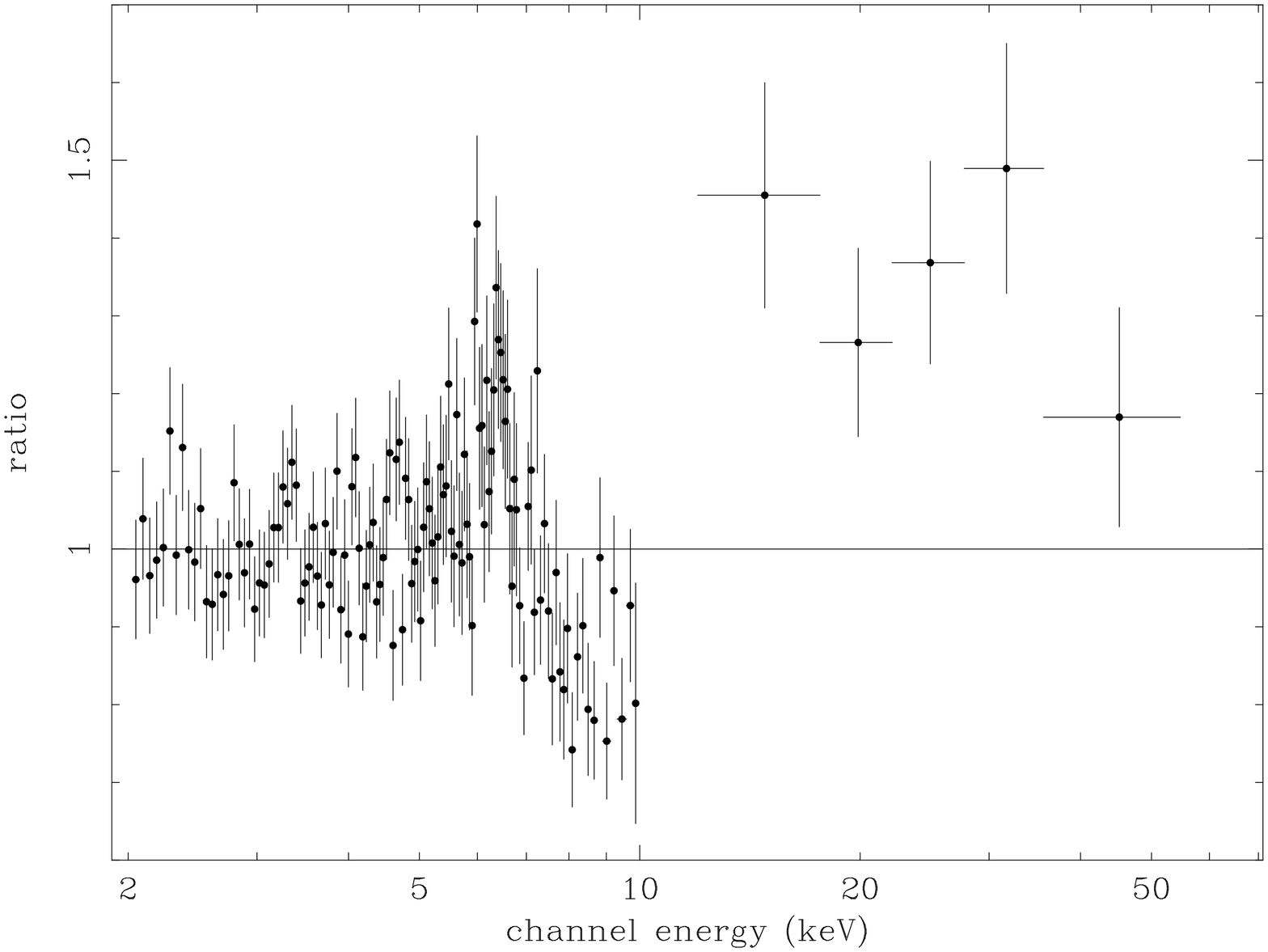,height=6cm,width=8cm,angle=270}
\end{minipage}\hfill
\begin{minipage}[b]{.46\linewidth}
\psfig{figure=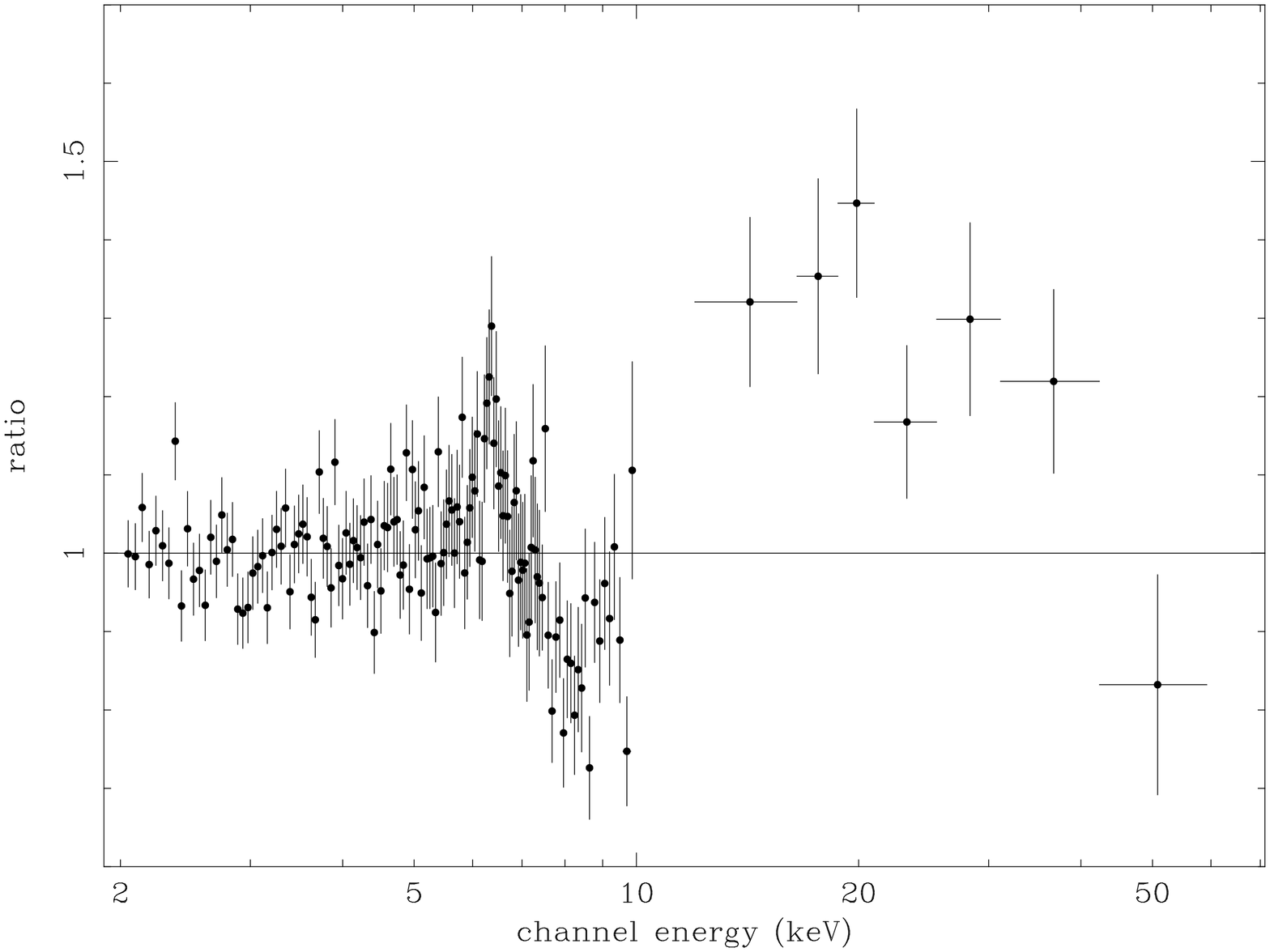,height=6cm,width=8cm,angle=270}
\end{minipage}\hfill
\caption{The MECS and PDS spectra of NGC 3516 obtained in 1996 November (left) 
and 1997 March (right), each divided by a best-fit single power-law. The photon 
indices are $\Gamma = 1.56 \pm 0.05$ and $\Gamma = 1.59 \pm 0.03$ respectively}
\end{figure*}

Within the BeppoSAX range, NGC 3516 has useful continuum emission from 0.2 to 60 
keV. The spectra of both epochs are partly shown in Fig.1, each compared to a 
best-fit power-law. The photon indices obtained for this simple model are in 
agreement with those yielded by the ASCA data in \cite{Kriss96} and by part of 
the Ginga data in \cite{Kolman93}. The spectra display features 
characteristic of Seyfert~1 galaxies in this band: the Fe K$\alpha$ line is 
clearly visible, and a reflection hump is present between 13 and 40--50 keV. To 
our knowledge, this is the first detection of a reflection hump in NGC 3516. 

At the second epoch the source was stronger by a factor 2 with respect to the 
first observation. Weak variability ($\sim 20$\%) was detected on timescales of 
$10^4$~s.

\section{THE FE K$\alpha$ LINE AND EDGE}
\begin{figure*}
\begin{minipage}[b]{.46\linewidth}
\psfig{figure=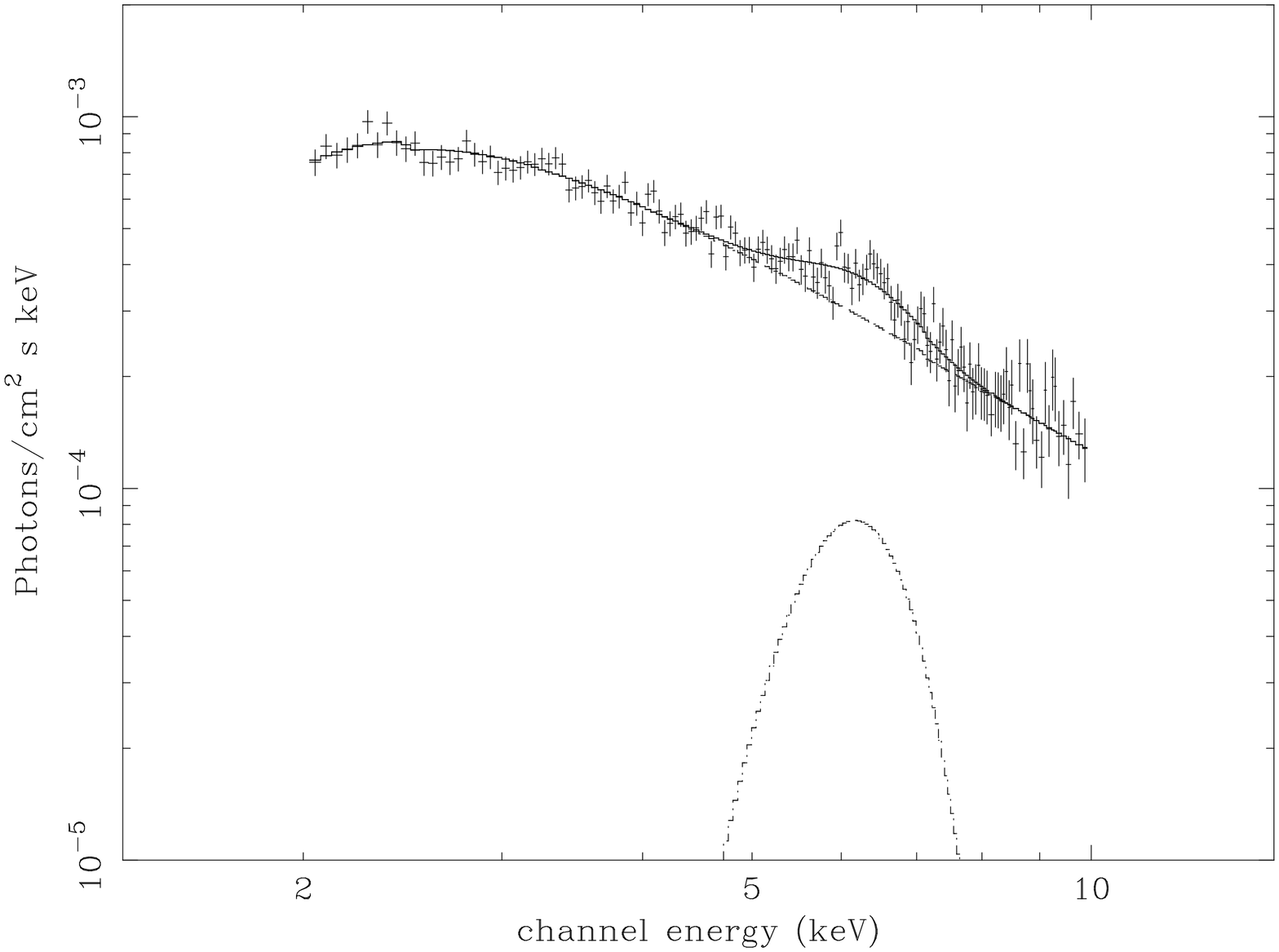,height=6cm,width=8cm,angle=270}
\end{minipage}\hfill
\begin{minipage}[b]{.46\linewidth}
\psfig{figure=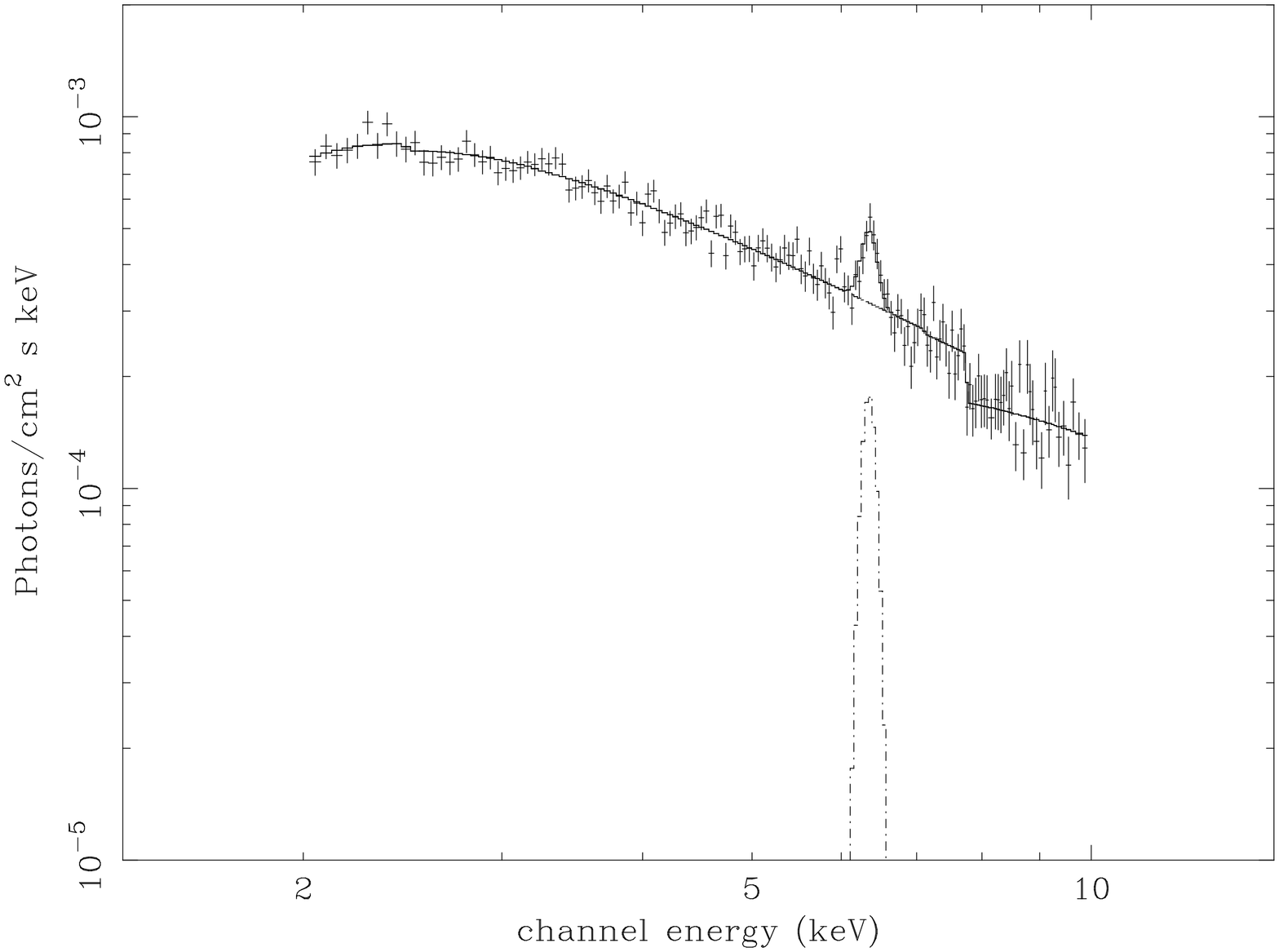,height=6cm,width=8cm,angle=270}
\end{minipage}\hfill
\caption{The MECS spectrum of 1996 November, fitted with a power-law and two 
different models for the Fe K$\alpha$ profile: a broad profile 
(left), and a narrow line + Fe edge (right).}
\end{figure*}

The Fe K$\alpha$ emission line
is well visible in both spectra.  Although there is an indication that the
profile is red-asymmetric, there are not sufficient constraints to discriminate
strongly between a broad accretion-disk profile, and a narrow emission line + Fe 
edge
model (both shown in Fig.2, fitted to the 1996 spectrum).  The results of the 
fits of both models to our spectra are listed in Table~1. The parameters of 
either model do not differ significantly at the two epochs of observation. The 
narrow-line + edge model
gives a 
slightly more significant fit ($\chi^2=154.4$ on 160 d.o.f. against 
$\chi^2=163.1$ 
on 161 d.o.f. in the 1996 spectrum).
In view of this, and because the EW and width in the broad line model seem 
unreasonably large, we believe that the narrow-line model is more acceptable. 
While our results for the narrow line and edge energies agree with those from 
the warm absorber fits of \cite{Kolman93} to Ginga data, our equivalent width 
values are significantly lower. The 
present results for the iron line EW
are also in agreement with the ASCA results \cite{Kriss96}. 
A detailed comparison is not straightforward, as in \cite{Kriss96} 
the iron line has been modelled with a broad plus a narrow component.
The lower resolution of BeppoSAX does not constrain this model
sufficiently.

\begin{table*}[hbt]
\setlength{\tabcolsep}{1.5pc}
\newlength{\digitwidth} \settowidth{\digitwidth}{\rm 0}
\catcode`?=\active \def?{\kern\digitwidth}
\caption{Results of fits to Fe line and edge}
\begin{tabular*}{\textwidth}{@{}l@{\extracolsep{\fill}}llll}
\hline
                 & \multicolumn{3}{l}{Broad line model} \\
\cline{2-4}
                 & \multicolumn{1}{l}{E (keV)}
                 & \multicolumn{1}{l}{$\sigma$ (keV)}
                 & \multicolumn{1}{l}{EW (eV)} \\
\cline{1-4}
1996 Nov. & \multicolumn{1}{l}{$6.21\pm0.20$} & $0.73^{+0.29}_{-0.25}$ 
          & $500^{+221}_{-157}$ \\
1997 Mar. & \multicolumn{1}{l}{$6.05^{+0.30}_{-0.45}$} & $0.97^{+0.48}_{-0.35}$ 
          & $464^{+329}_{-175}$ \\
\hline
                 & \multicolumn{4}{l}{Narrow line + edge model} \\
\cline{2-5}
                 & \multicolumn{2}{l}{Line}
                 & \multicolumn{2}{l}{Edge} \\
\cline{2-3} \cline{4-5}
                 & \multicolumn{1}{l}{E (keV)}
                 & \multicolumn{1}{l}{EW (eV)}
                 & \multicolumn{1}{l}{E (keV)}
                 & \multicolumn{1}{l}{$\tau$} \\
\hline
1996 Nov. & \multicolumn{1}{l}{$6.41\pm0.09$} & $142\pm46$ 
          & $7.82^{+0.25}_{-0.22}$? & $0.30^{+0.12}_{-0.10}$? \\
1997 Mar. & \multicolumn{1}{l}{$6.45\pm0.10$} & ?$86\pm35$ & $7.75\pm0.16$ 
          & $0.28\pm0.08$ \\
\hline
\end{tabular*}
\end{table*}

\section{THE WARM ABSORBER}

\begin{figure*}
\begin{minipage}[b]{.46\linewidth}
\psfig{figure=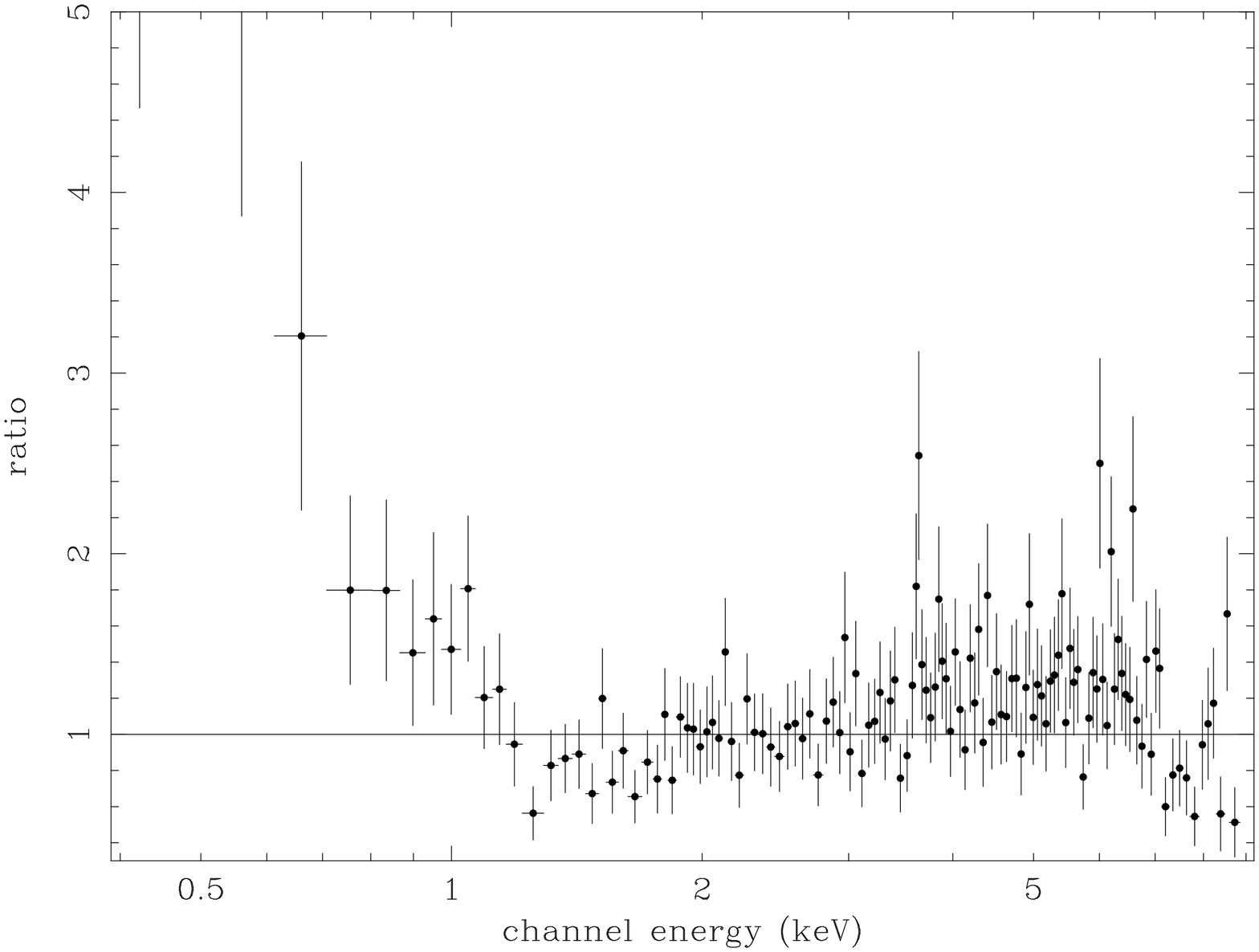,height=6cm,width=8cm,angle=270}
\end{minipage}\hfill
\begin{minipage}[b]{.46\linewidth}
\psfig{figure=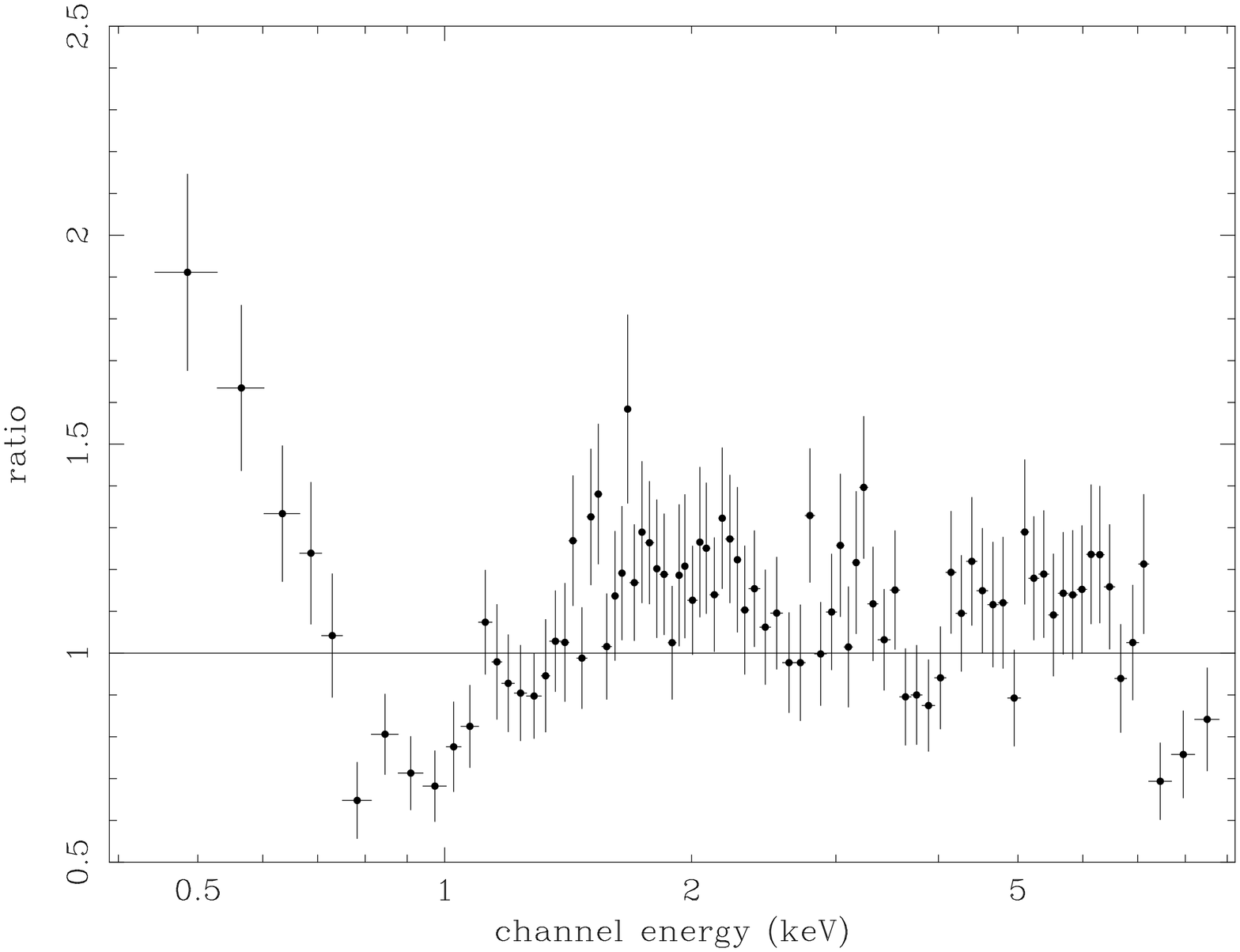,height=6cm,width=8cm,angle=270}
\end{minipage}\hfill
\caption{The LECS spectra of 1996 November (left) and 1997 March (right) divided 
by a power-law. Note that no absorption edge is visible at $\sim0.8$~keV in the 
first spectrum, while it is clearly present in the later observation.}
\end{figure*}

One of the most surprising results of these observations concerns the absorption 
edge at $\sim 0.8$~keV. While it is clearly visible in the second, stronger 
spectrum, there is no evidence of its presence in the first one (Fig.3). 
There 
is thus 
evidence for a complex ionization structure, which almost certainly depends on 
the continuum flux, as well as possibly on time. While the XUV model proposed in 
\cite{Mathur97} cannot be confirmed or disproved at this stage, it will have to 
take into account the effects of flux variations on time-scales of a few months.

A preliminary fit to the absorption in the 1997 spectrum yields $\tau_{\rm 
OVII}=1.27^{+0.53}_{-0.40}$ and $\tau_{\rm OVIII} < 0.68$, which are in formal 
agreement with most of the fits to ROSAT data performed by \cite{Mathur97}.

\section{CONCLUSIONS}
We have presented preliminary results from two BeppoSAX observations of NGC 
3516. Useful emission was detected up to 60~keV, which gives unprecedented 
wavelength coverage of the X-ray spectrum of this bright Seyfert 1 galaxy. 
Several features typical of Seyferts --- Fe K$\alpha$ line, reflection hump at 
high energies, and a warm absorber edge at $\sim0.8$~keV in the later spectrum 
--- are visible and will be the object of further study. Of particular note is 
the strong variation of the absorption edge, which was not detected in the 
first, weaker spectrum. This will require that current XUV 
absorber models take into account the effects of flux variations on time-scales 
shorter than those required for the dynamical evolution of the absorbing gas. In 
any case, the fact that the Fe edge is visible in both spectra while the 
O~VII/O~VIII edge is not implies that they are not produced by the same gas. 

\bigskip

This project has received partial financial support from the Italian Space 
Agency (ASI contract ARS--96--70). We are extremely grateful to the BeppoSAX 
team for their continuous and generous support.

\end{document}